\def\twocol{2}

\def\colopt{\twocol}
\ifnum\colopt=\twocol
 \documentstyle[floats,twocolumn,aps,psfig]{revtex}
 \def\Section#1{}
\else
 \documentstyle[floats,preprint,aps,psfig]{revtex}
 \def\Section#1{\section{#1}}
\fi
\def\EQ{\begin{equation}}
\def\EN{\end{equation}}
\def\bea{\begin{eqnarray}}
\def\eea{\end{eqnarray}}

\def\D{\Delta}
\def\no{\noindent}

\begin{document}

\ifnum\colopt=\twocol
 \twocolumn[\hsize\textwidth\columnwidth\hsize\csname @twocolumnfalse\endcsname
\fi

\title{Universal Ratios in the 2-D Tricritical Ising Model}
\author{D. Fioravanti$^{1,2}$, G. Mussardo$^{2,3}$ and P. Simon$^{1,2}$
}
\address{
$^{1}$International School for Advanced Studies,
     Via Beirut 2-4, 34014 Trieste, Italy.\\
$^{2}$Instituto Nazionale di Fisica Nucleare, Sezione di Trieste\\
$^{3}$Dipartimento di Fisica, Universita dell'Insubria, Como, Italy\\
}

\maketitle
\begin{abstract}
\begin{center}
\parbox{14cm}{We consider the universality class of the 
two--dimensional Tricritical Ising Model. The scaling form 
of the free--energy naturally leads to the definition of  
universal ratios of critical amplitudes which may have experimental 
relevance. We compute these universal ratios by a combined use of 
results coming from Perturbed Conformal Field Theory, Integrable 
Quantum Field Theory and numerical methods.}  
\end{center}
\end{abstract}

\ifnum\colopt=\twocol
 \vskip1pc]
\fi

\vspace{.3cm}

An unifying principle in the study of critical phenomena goes 
under the name of {\em universality} \cite{universality}. In the 
vicinity of a  phase transition, when the correlation length is 
much larger than any microscopic scale, one can assign each system 
to a universality class, which is identified by its dimensionality 
$D$, the symmetry properties of the order parameters and the number 
of relevant fields. The first characteristic of a given universality 
class is the set of critical exponents, expressed in terms 
of algebraic expressions of the conformal dimensions of the relevant 
fields. Additional data of a universality class may be derived by 
the scaling properties of the free--energy alone. These data -- 
called {\em universal ratios} -- are pure numbers, obtained by 
taking particular combinations of various thermodynamical amplitudes 
in such a way to cancel any dependence on the microscopic scales. 
Together with critical exponents, universal ratios are ideal 
fingerprints of the universality classes. From an experimental 
point of view, there is by now a large literature on universal 
ratio measurements of various systems extending from binary 
fluids to magnetic systems and polymer conformations (for an 
extensive review on the subject, see \cite{Privman}).

In recent years, due to the theoretical progress achieved in the 
study of two--dimensional models (at criticality by the 
methods of Conformal Field Theory (CFT) \cite{BPZ}, and 
away from criticality by the approach of Perturbed Conformal 
Theories \cite{Zamaway}), several universal quantities have 
been computed by different techniques for a large variety of 
bidimensional systems, such  as the self--avoiding walks 
\cite{pol,CMpol}, 
the Ising model  \cite{DMIMMF,Delfino,Sokal,Cas}, the q-state 
Potts model \cite{Potts}, to name few. In this letter we will focus 
on the first determination of some universal ratios relative 
to the class of universality of the $2$--$D$ Tricritical Ising 
Model (TIM) for which very few universal quantities are known 
(see \cite{Privman,Lawrie}). Whereas the $3$--$D$ TIM describes, for
instance, the universality class of an anti--ferromagnet with strong 
uniaxial anisotropy like FeCl$_2$, its $2D$ version can describe 
the tricritical behaviour of a binary mixture of thin films of 
$He^3-He^4$ \cite{Berker} or order--disorder transitions in absorbed 
systems \cite{Tejwani} (for a review on the theory of tricritical 
points, see \cite{Lawrie}). Hence there is an obvious interest in 
computing the amplest set of universal data for this universality 
class and in testing the theoretical predictions versus their 
experimental determinations.

In a continuum version of the TIM (which is, after all, a particular 
representative of this universality class), it is convenient 
to adopt a Landau--Ginzburg (LG) formulation based on a scalar field 
$\Phi(x)$ with $\Phi^6$ interaction. The LG approach permits to have a 
clear bookkeeping of the symmetry properties of each order parameter 
and to easily understand the phase diagram of the model, at least
qualitatively. The class of universality of the TIM is then described 
by the LG euclidean action 
\[
{\cal A} =  \int d^Dx \left[\frac{1}{2} (\partial_{\mu} \Phi)^2 + 
g_1 \Phi + g_2 \Phi^2 + g_3 \Phi^3 + g_4 \Phi^4 + 
\Phi^6 \right]
\]
with the tricritical point identified by the bare conditions 
$g_1=g_2=g_3=g_4=0$. Adopting a magnetic terminology, the 
statistical interpretation of the coupling constants is 
as follows: $g_1$ plays the role of an external magnetic 
field $h$, $g_2$ measures the displacement of the temperature from 
its critical value $(T-T_c)$, $g_3$ may be regarded as 
a staggered magnetic field $h'$ and finally $g_4$ may be 
thought as a chemical potential $\mu$ for the vacancy density. 
Dimensional analysis shows that the upper critical dimension of 
the model is $D=3$, where tricritical exponents are expected  
to have their classical values (apart logarithmic corrections). 
In two dimensions, although the mean field solution of the model 
cannot be trusted for the strong fluctuations of the order parameters,   
an exact solution at criticality is provided by CFT. In fact, 
the TIM is described by the second model of the unitary minimal 
series of CFT \cite{BPZ}, with central charge equal to $C = 
\frac{7}{10}$. There are six primary fields, identified 
with the normal ordered composite LG fields \cite{ZamLG}, which 
close an algebra under the Operator Product Expansion (OPE). 
Only four of them are relevant ({\it i.e.} with conformal 
dimension $\Delta < 1$): $\sigma = \varphi_1\equiv\Phi$ ($\Delta_1 
= \frac{3}{80}$), $\varepsilon = \varphi_2 \equiv :\Phi^2:$ 
($\Delta_2 = \frac{1}{10}$), $\sigma' = \varphi_3 \equiv  
:\Phi^3:$ ($\Delta_3=\frac{7}{16}$) and $t = \varphi_4\equiv 
:\Phi^4:$ ($\Delta_4 = \frac{3}{5}$). The fields $\varepsilon$ 
and $t$ are even under the $Z_2$ spin--symmetry whereas 
$\sigma$ and $\sigma'$ are odd. There is another $Z_2$ symmetry 
of the model (related to its self-duality), under which 
$D \varepsilon D^{-1} = -\varepsilon$, $D t D^{-1} = t$, whereas 
the magnetic order parameters are mapped onto their corresponding  
disorder parameters. Each of the above relevant fields can be 
used to move the TIM away from criticality (the resulting phases 
of the model are discussed in \cite{LMC}).  

In order to derive the scaling form of the free energy and the set
of universal ratios for the $2$--$D$ TIM, let us first normalise 
the two-point functions of the fields as $\langle \varphi_i 
(r) \varphi_i(0) \rangle \sim \frac{A_i}{r^{4 \Delta_i}}$ when $r 
\to 0$ (in the perturbed CFT approach to the model, $A_i =1$). When 
the TIM is moved away from criticality by means of one (or several) 
of its relevant fields, with the resulting action 
${\cal A} = {\cal A}_{CFT} + \sum_p g_p \int d^2x\,\varphi_p(x)$, 
a finite correlation length $\xi$ generally appears. Its 
scaling form may be written in four possible equivalent ways, 
according to which coupling constant is selected out as 
a prefactor
\EQ
\xi = a \,(K_i g_i)^
{-\frac{1}{2 -2 \Delta_i}}\, 
{\cal L}_i\left(\frac{K_j g_j}{(K_i g_i)^{\phi_{ji}}}
\right) \,\,\,, 
\label{xii}
\EN 
where $a$ is some microscopic length scale, $\phi_{ji} \equiv 
\frac{1- \Delta_j}{1- \Delta_i}$, and ${\cal L}_i$ are 
universal homegeneous scaling functions of the ratios 
$\frac{K_j g_j}{(K_i g_i)^{\phi_{ji}}}$. The terms $K_i$ are 
non--universal metric factors which depend on the unit chosen 
for measuring the external source $g_i$, alias on the particular 
realization of the universality class. Let $f[g_1,g_2,g_3,g_4]$ be 
the singular part of the free--energy (per unit volume). According 
to which coupling constant is selected out as a prefactor, it can be 
parameterised in four possible equivalent ways as: 
\EQ
f[g_1,g_2,g_3,g_4]  \equiv 
\left(K_i g_i\right)^{-\frac{1}{1-\Delta_i}} \,
{\cal F}_i\left(\frac{K_j g_j}{(K_i g_i)^{\phi_{ji}}}\right) \,, 
\label{scalingfree}
\EN 
where ${\cal F}_i$ are scaling functions. For the Vacuum Expectation 
Value (VEV) of the fields $\varphi_j$ in the $i^{\rm th}$ direction 
({\it i.e.} for the off--critical theory finally obtained by $g_i \neq 0$,  
$g_k=0$, $k \neq i$), we have  
\EQ
\langle \varphi_j \rangle_i = -\frac{\partial f}{\partial g_j} 
\left|_{g_k=0} \equiv B_{ji} 
g_i^{\frac{\Delta_j}{1- \Delta_i}} \right.\,\,\,,
\label{bji}
\EN 
where, from (\ref{scalingfree}), $B_{ji} \sim K_j 
K_i^{\frac{\Delta_j}{1-\Delta_i}}$. In  a similar manner, 
for the generalized susceptibilities we have  
\EQ
\hat \Gamma_{jl}^i =   
-\frac{\partial^2 f}{\partial g_l \partial g_j}\left|_{g_k=0}
\equiv \Gamma_{jl}^i \,g_i^
{\frac{ \Delta_j +  \Delta_l - 1}
{1 -  \Delta_i}} \,\,\, ,\right.
\label{susc}
\EN
where, from (\ref{scalingfree}), $\Gamma_{jl}^i \sim K_j K_l 
K_i^{\frac{ \Delta_j +  \Delta_l - 1}{1 -  \Delta_i}}$. These 
quantities are obviously symmetric in the lower indices 
($\hat \Gamma^{i}_{22}$ and $\hat \Gamma^{i}_{11}$ are respectively
the usual specific heat and magnetic susceptibility in the $i^{\rm th}$ 
direction). Similarly, for the correlation length we have 
$\xi_i = a \,\xi_0 \,g_i^{-\frac{1}{2-2\Delta_i}}$, with 
$\xi_0 \sim K_i^{-\frac{1}{2-2\Delta_i}}$. From the above formulas, 
appropriate combinations can be found such that the non--universal 
metric factors $K_i$ cancel out. Some of the $2$--$D$ universal 
ratios are: 
\bea
(R_c)^i_{jk} &=& \frac{\Gamma_{ii}^i \Gamma_{jk}^i}{B_{ji} 
B_{ki}}
\label{Rc}\\
(R_{\chi})^i_j  &=& \Gamma_{jj}^i B_{jj}^{ {\Delta_j - 1
\over \Delta_j}}
B_{ji}^ {\frac{1-2 \Delta_j}{ \Delta_j}} 
\label{Rchi}\\
R^i_{\xi}  &=& \left(\Gamma_{ii}^i\right)^{1/2} 
\xi_i^0 
\label{Rxi}\\
(R_A)^i_j  &=& \Gamma_{jj}^i \, B_{ii}^
{\frac{-2 \Delta_j  -  
\Delta_i +2}{ \Delta_i}} \, 
B_{ij}^{\frac{2 \Delta_j -2}{\Delta_i}} 
\label{RA}\\
(Q_2)^i_{jk}  &=& \frac{\Gamma^i_{jj}}
{\Gamma^k_{jj}} 
\left(\frac{\xi_k^0}{\xi_j^0}\right)^{2-4 
\Delta_j} \,\,\,.
\label{Q2}
\eea
In this letter we only consider the case $i=1,2$, which 
correspond to the most important physical deformations of the  
model (the magnetic and the thermal ones), {\it i.e.} those 
which are most accessible from an experimental point of view. 
For both the magnetic and thermal deformation there are no 
mixing among the conformal fields due to ultraviolet 
renormalization \cite{ZamYL,DSC}. A complete analysis 
relative to all deformations of the TIM and the theoretical 
details of our approach will be published elsewhere \cite{FMS}. 

The $\epsilon$ perturbation around the critical TIM is 
integrable and its behavior is governed by the $E_7$ 
algebra \cite{MC}. Therefore the $B_{j2}$'s in eq.\,(\ref{bji}) 
have been computed exactly in \cite{vacuum}. On the 
other hand, the $\sigma$ perturbation is non-integrable 
(numerical indications were discussed in \cite{LMC}). 
In this case, the $B_{j1}$'s have been numerically evaluated  
in \cite{GM1} by using the so--called Truncated Conformal 
Space Approach (TCSA) \cite{YZ}. This method consists in 
diagonalizing the off--critical Hamiltonian on a cylinder in 
a truncated conformal basis of the critical TIM such that 
an extimation of $\langle \varphi_j \rangle_1$ can be 
obtained from the knowledge of the eigenvectors (only the 
ground state eigenvector is needed for the VEV). All these 
calculations can be easily performed by means of the numerical 
program of ref.\,\cite{LM}. 

In order to estimate the universal ratios, it is still necessary  
to calculate the $\Gamma_{jk}^i$'s. Their values can be extracted 
in two different ways. The first method is purely numerical and 
of immediate use, since it consists in employing the TCSA to compute 
numerically the derivative  $\frac{\partial}{\partial g_k} \langle
\varphi_j\rangle_i$ (details will be found in \cite{FMS}). The second 
method is based on the fluctuation--dissipation theorem which 
permits to express the generalised susceptibilities as  
\EQ
\hat\Gamma_{jk}^i = \int d^2 x \,\langle \varphi_j(x) 
\varphi_k\rangle_{i}^c
\, ,
\label{fluctdis}
\EN 
where $\langle \cdots\rangle^c$ indicates the connected correlator.
Therefore in this second approach we first need to evaluate the 
$2-$point correlation functions and then to perform the integration. 
For our calculation of the universal ratios, we have employed  
both methods, finding an agreement in their final outputs. Let 
us briefly discuss the second method. First of all, write the 
integral (\ref{fluctdis}) in polar coordinate as $\hat\Gamma_{jk}^i 
= 2\pi \int dr \,r \,\langle \varphi_j(r) \varphi_k\rangle_i^c$. 
Secondly, decompose the integral over $r$ into two integrals 
over the regions $0 < r < R$ and $r \geq R$ with $R\sim\xi$. 
When $r< R$, the correlation function $\langle \varphi_j(r) 
\varphi_k(0)\rangle_{i}$ can be efficiently evaluated by 
using a short--distance expansion \cite{ZamYL}
\EQ
\langle \varphi_j(r) \varphi_k(0)\rangle_i
= \sum_l C_{jk}^l(r)\langle \varphi_l\rangle_i 
\label{OPE}
\EN 
where the non--analytic dependence on the coupling constant is 
completly encoded into the VEV's, whereas the structure constants 
$C_{jk}^l(r)$ can be evaluated perturbatively in $g$ 
\EQ
C_{jk}^l(r) = r^{2 (\Delta_l - \Delta_j -\Delta_k)} \,
\sum_{n=0}^{\infty} C_{jk}^{l (n)} \left( 
g_i r^{2-2 \Delta_i}\right)^n  \,\,\,.
\label{perturbativeOPE}
\EN 
For the TIM, $C_{jk}^{l (0)}$ have been computed in \cite{LMC} whereas 
their first correction can be obtained by the formula 
\EQ
C_{jk}^{l(1)} = - \int ^{'} d^2 z~  \langle
\varphi_l(\infty)\varphi_i(z)\varphi_k(1)\varphi_j(0)\rangle_{CFT}
\,\,\, ,
\label{CFTcorrection}
\EN 
where the prime indicates a suitable infrared regularization of 
the integral. As shown in \cite{GM2}, an efficient way to compute 
the regularised integrals is through a Mellin transformation. 
Hence, the calculation of the above integral (\ref{CFTcorrection}) 
on the conformal functions plus the knowledge of the various 
expectation values $\langle \varphi_l\rangle_i $ enables us 
to reach a quite accurate approximation of $\langle \varphi_j(r) 
\varphi_k(0)\rangle_i$ in the ultraviolet limit, {\it i.e.} for 
$r < R$. By choosing $R\sim \xi$, one can obtain an overlap 
between the ultraviolet and the infrared representations of the 
correlation functions. The latter is expressed by means of the 
spectral series of the correlators on the massive states 
$\mid A_{k}(\theta)\rangle$ of the off--critical theory 
\EQ
\langle\varphi_j(x) \varphi_k(0)\rangle_c = \sum_{n=1}^{\infty}
g_n(r) \,\,\,,
\label{g_n}
\EN
where
\begin{eqnarray*}
& & g_n(r)  =  \int
\frac{d\theta_1}{2\pi} \cdots \frac{d\theta_n}{2\pi}\,
\langle 0|\varphi_j(0)|
A_{a_1}(\theta_1) \cdots A_{a_n}(\theta_n)
\rangle  \times \\
& &  \,\,
\langle A_{a_1}(\theta_1) \cdots A_{a_n}(\theta_n)|
\varphi_k(0)|0 \rangle \,e^{-r \sum_{k=1}^n m_k \cosh\theta_k}
\,\,\,.
\nonumber
\end{eqnarray*} 
As tested in several examples (see, for instance
\cite{CMpol,DMIMMF,ZamYL,GM2,DMM35}), the above series (\ref{g_n}) 
converges very fast even for $r \sim \xi$ so that its 
truncation to the lowest terms is able to capture the correct 
behaviour of the correlator in the interval $r \geq \xi$. For 
the integrable theory defined by the thermal deformation 
of the TIM, one can truncated the series up to the lowest 
$2$--particle states, with the relative matrix elements  
computed along the lines of the refs.\, 
\cite{DMIMMF,ZamYL,AMV,Smirnov}. For the non--integrable theory 
defined by the magnetic deformation of the TIM, it is hard to go 
beyond the one--particle matrix elements and one has to be satisfied 
with the estimate of the correlators obtained by the one--particle 
contributions only: since this theory has two lowest masses with 
mass ratio $m_2 \simeq 2 m_1 \cos\frac{\pi}{5}$ \cite{LMC}, in 
this case we have 
\[
\langle \varphi_j(r) \varphi_k(0)\rangle_i \approx 
{1\over \pi}\left(f_j^1 f_k^1 K_0(m_1r) +f_j^2 f_k^2 
K_0(m_2r)\right)
\]
where $K_0(x)$ is the modified Bessel function and the indices 
$1,2$ refer to the first and second massive states. The one--particle 
matrix elements of this model $f_j^k = \langle 0 |\varphi_j(0)|A_k 
\rangle$ can be also computed numerically by using the TCSA 
\cite{FMS}. 

Once an overlap of the short and large distance expansions  
of the correlators in the region $r\sim\xi$ has been checked, 
a numerical integration of the correlators provides the 
$\Gamma_{jk}^i$'s. An explicit test of the validity of the 
above method (with a corresponding estimate of its errors) 
is provided by the comparison of the values of $\Gamma_{ik}^i$ 
(obtained by the numerical integration) with their exact 
determination extracted by the $\Delta-$theorem sum rule, 
when this theorem applies \cite{DSC}: 
\EQ
\Gamma_{ik}^i=-{\D_k\over 1-\D_k} B_{ki} \,\,\,.
\EN
This check shows that the uncertanties for $\Gamma_{jk}^i$ is 
at worst about $5\%$, better for the strongest relevant operators. 
Gathering all these results, a set of universal ratios for the 
TIM have been obtained. Some of them are exact, like $(R_c)^{1}_{1,k}
= \frac{240}{5929} \Delta_k$, $(R_c)^2_{2,k} = \frac{10}{81} 
\Delta_k$ ($k=1,\ldots,4$). We have also computed those relative 
to the low and high temperature phase of the model (Table 1). 
An interesting universal ratio is provided in this case by the 
correlation length prefactors $\xi_0^{\mp}$, below and above the 
critical temperature (as extracted from the correlation function 
of the magnetic operator using its duality properties)
\EQ
{\xi_0^-\over \xi_0^+}= 2\cos\left({5\pi\over 18}\right) \approx 
1.28557...
\label{ratxi0}
\EN
which can be inferred by the exact mass spectrum of the model and the 
parity properties of the excitations \cite{LMC,MC}.

In summary, we have combined techniques coming from CFT, integrable 
models and numerical methods to obtain for the first time a set of 
universal quantities for the class of universality of the $2D$ 
Tricritical Ising Model. It would be interesting to have an 
experimental determination of these quantities and a comparison 
with the theoretical predictions presented here. 

This work done under partial support of the EC TMR Programme
{\em Integrability, non-per\-turba\-tive effects and symmetry in
Quantum Field Theories}. We would like also to thank A.B. and 
Al.B. Zamolodchikov for useful discussions. We are also grateful 
to V. Rittenberg and M. den Nijs for suggestions and for their 
interest in this work.

\vspace{15mm}

\noindent {\bf Table 1:} Amplitude ratios 
$R^2_{jk}={\Gamma_{jk}^{2+}\over \Gamma_{jk}^{2-}}$.
\vskip 3mm
\begin{center}
\begin{tabular}{|ccc||ccc|}
\hline
$R^2_{11} $   &  = &$3.54$  &
$R^2_{13} $   &  = &$-2.06 $  \\
\hline
$R_{22}^2 $   &  = &$1$  &
$R_{24}^2 $   &  = &$-1$  \\
\hline
$R_{33}^2 $   & = &$ 1.30$  &
$R_{44}^2$   & = &$1$  \\
\hline
\end{tabular}
\end{center}

\no {\bf Table 2:} Universal ratios $(R_c)_{jk}^{1}$ and $(R_c)_{jk}^{2-}$.
\vskip 3mm
\begin{center}
\begin{tabular}{|ccl||ccl|} \hline
$(R_c)_{22}^1$   & = & $ 1.05~10^{-2}$   &  $(R_c)_{23}^1$    
& = & $4.85~10^{-2}$  \\
\hline
$(R_c)_{24}^1$   & = & $ 6.7~10^{-2} $   &  $(R_c)_{33}^1$    
& = & $3.8~10^{-1} $ \\
\hline \hline
$(R_c)_{11}^{2-}$ & = &$ 2.0~10^{-3}$  & $(R_c)_{14}^{2-}$ & = & 
$-2.34~10^{-2}$  \\
\hline
$(R_c)_{13}^{2-}$ & = &$ 1.79~10^{-2}$ & $(R_c)_{33}^{2-}$ &  = & 
$3.4~10^{-1}$  \\
\hline
\end{tabular}
\end{center}

\no {\bf Table 3:} Universal ratio $(R_{\chi})_{j}^{i}$ for $i,j=1,2$.
\vskip 3mm
\begin{center}
\begin{tabular}{|ccl||ccl|} \hline
$(R_{\chi})_{1}^1$   & = &  $3.897~10^{-2}$ & $(R_{\chi})_2^{2+} 
$  & = & $0.1111$  \\
\hline
$(R_{\chi})_{2}^1$   & = &  $0.116$         & $(R_{\chi})_{1}^{2-} 
$  & = & $0.040$  \\
\hline
$(R_{\chi})_{1}^{2+}$ & = & $ 0 $           & $(R_{\chi})_{2}^{2-} 
$ & =  & $0.1111$  \\
\hline
\end{tabular}
\end{center}

\no {\bf Table 4:} Universal ratios $R_{\xi}^i$ and 
$(R_A)_{j}^{i}$ for $i,j=1,2^-,2^+$.
\vskip 3mm
\begin{center}
\begin{tabular}{|ccl||ccl|} \hline
$R_{\xi}^1$      & = & $ 7.557~10^{-2}$     &  & & \\
\hline
$R_{\xi}^{2+}$   & = & $ 1.0784~10^{-1} $   &  $R_{\xi}^{2-}$    
& = & $8.389~10^{-1} $ \\
\hline \hline
$(R_A)_{2+}^{1}$ & = & $0 $  & $(R_A)_{2-}^{1}$ & = & 
$3.918~10^{-2}$  \\
\hline
$(R_A)^{2+}_{1}$ & = & $ 2.958~10^{-1}$ & $(R_A)^{2-}_{1}$ &  = & 
$8.260~10^{-1}$  \\
\hline
\end{tabular}
\end{center}

\no {\bf Table 5:} Universal ratios $(Q_2)^i_{jk}$ for $i,j,k=1,2^+,2^-$.
\vskip 3mm
\begin{center}
\begin{tabular}{|ccl||ccl|} \hline
$(Q_2)^1_{2^+1}$   & = & $ 1.260$     & $(Q_2)^1_{2^-1}$  
& = &  $1.884$ \\
\hline
$(Q_2)^1_{2^+2^+}$ & = & $ 1.973$     & $(Q_2)^1_{2^+2^-}$     
& = &  $1.320$ \\
\hline 
$(Q_2)^{2+}_{11}$  & = & $ 1.56 $    & $(Q_2)^{2-}_{11}$ 
& = &  $0.442$  \\
\hline
$(Q_2)^{2+}_{12^-}$ & = & $ 1.70$     &  &   & \\
\hline
\end{tabular}
\end{center}
\end{document}